\documentstyle[epsfig]{elsart}

\begin{document}

\begin{frontmatter}

\title{The mean-field theory for attraction between 
like-charged macromolecules}

\author{Jeferson J. Arenzon, Yan Levin}
\footnote{Corresponding author levin@if.ufrgs.br}  

\address{Instituto de F\'{\i}sica -- UFRGS \\
CP 15051 -- 91501-970 -- Porto Alegre RS -- Brazil}

\author{J\"urgen F. Stilck}

\address{Instituto de F\'{\i}sica -- UFF \\
Av. Litor{\^a}nea, s/n$^o$, 24210-340, Niter{\'o}i, RJ, Brazil}

\begin{abstract}

A  mean-field theory based on Gibbs-Bogoliubov inequality
is constructed to study the interactions between two like-charged
polyions.  It is shown that contrary to the previously established
paradigm, a properly constructed mean-field theory can 
quantitatively account for the attractive interactions between
two like-charged rods.

\end{abstract}

\end{frontmatter}

One of the most fascinating problems that has recently appeared
in the field of condensed matter physics is the discovery of attraction
between like-charged macromolecules\cite{Oos68}.  
This attraction plays a fundamental 
role in various biological processes such as the condensation of 
DNA \cite{Blo91,Pod94}
and the formation of fibers composing cellular cytosceleton \cite{Tan96}.  The
attraction between like-charged colloids has also been observed in 
various experiments and simulations \cite{Pat80,Jen97,All98,Jo96}.  
It has been noted 
that the attraction appears only in the presence of multivalent
counterions.

A number of models have been proposed to try to
explain the mechanism of these strange phenomena.  It is now clear,
from both  simulations and  experiments,  that this effect
is purely electrostatic and is produced by  strong many 
body interactions present in polyelectrolyte solutions. In a
beautiful set of experiments Tang {\em et al}\cite{Tan97} demonstrated 
how  addition
of simple monovalent salt produced  dissociation of the actin
bundles.  The F-actin chains are highly charged polymers,  
which inspite of their large negative charge density, aggregate in well
defined bundles in the presence of polyamines. However, this bundling
can be reversed by addition of simple monovalent salt which
screens the electrostatic interactions between the polyions and
the multivalent counterions.

The first explanation of attraction between like charged surfaces
in the presence of multivalent counterions was advanced by 
Kjellander and Marcelja \cite{Kje84} based on the integral 
equation formalism. From the 
numerical solution of the AHNC equation these authors came to conclude
that for sufficiently high surface charge,  an attraction
can arise between like charged plates. A very simple physical 
picture to explain the mechanism of 
attraction was advanced by Rouzina and Bloomfield \cite{Roz96}, and extended
by Shklovskii \cite{Shk98}.  
These authors proposed that the condensed counterions
around the two plates form strongly coupled  Wigner crystals.  
In the case of rod-like
polyions, a similar explanation has been advanced by Arenzon {\em et al}
on the basis of an exactly solvable model\cite{Are99,monica}.  A
different mechanism, relying on correlated fluctuations, has been proposed
by Ha and Liu \cite{Ha97}, but has been criticized by Levin {\em et al}
\cite{Lev99}.

Since the beginning of the study of this interesting phenomenon there
has been a general consensus that the attraction must arise as
a result of correlations of condensed counterions \cite{Ray94}.  
It was, therefore, implicitly assumed that
no mean-field theory would be able to account for this phnenomenon.  This
belief was further reinforced by the solutions of Poisson-Boltzmann
equation (PB) which, of course, 
did not predict any attraction.  Not all mean-fields, however, are
equal.  In this paper we shall present a mean-field theory, 
which quantitatively accounts  for the
attraction between like-charged rods in the presence of 
condensed multivalent
counterion.

We consider two parallel polyions modeled as rigid rods, 
each having  $Z$ charges of value  $-q$,
spaced uniformly with separation $b$ along the length.  The rods are 
separated by distance $d=xb$. The strong electrostatic
interaction between the polyions and the multivalent
counterions present in solution leads to counterion condensation
\cite{Man69,Lev96,Lev98,Kuh98}.
The effect of 
$n$, $\alpha$-valent condensed
counterions, is approximated by the renormalization of
local charge.  Thus, if one of the charged sites of a
polyion has an associated condensed counterion its effective charge
becomes $-q(1-\alpha)$. Note that in this simple
model the condensed counterions are assumed to reside only on top
of the charged sites. The net charge of each polyion is $(Z-\alpha n)q$. 
The Hamiltonian for the interactions between the two rods is \cite{Are99}, 
\begin{equation}
{\cal H}=\frac{1}{2D}\sum_{i,i'=1}^{Z} \sum_{m,m'=0}^{1}
\frac{q^2(1-\alpha\sigma_i^m)(1-\alpha\sigma_{i'}^{m'})}
{r(i,m;i',m')},
\end{equation}
where we have introduced  $m=0,1$ to label the two polyions. The
distance between two charged sites 
of the polyions, $(i,m) \neq (i',m')$ ,  is 
$r(i,m;i',m')=b\sqrt{|i-i'|^2 + (1-\delta_{mm'})x^2}$, and
 $\sigma_i^m$ is an occupation variable such that $\sigma_i^m=1$ if
the i'th site of m'th polyion has an associated counterion and
 $\sigma_i^m=0$ if this site is unoccupied.

The mean field theory can be constructed with the help of 
the Gibbs-Bogoliubov
bound for the free energy \cite{Kuh98}, 
$F\leq {\cal F}\equiv F_0 + <{\cal H} - {\cal H}_0>_0$.  The
average, $<\ldots >_0$, is performed with respect 
to the trial Hamiltonian ${\cal H}_0$. To make the calculation
as simple as possible we shall take this to be of one body
form,
\begin{equation}
\label{1}
{\cal H}_0=-q\sum_{i,m} (1-\alpha \sigma_i^m)\varphi_i^m \;,
\end{equation}
where $\varphi_i^m$ is the mean electrostatic potential experienced by
the $i$-$th$ monomer of the rod $m$. 
The upper bound for the  free energy can now be calculated,
\begin{eqnarray}
{\cal F} &=& \frac{1}{2D}\sum_{i,i'=1}^{Z} \sum_{m,m'=0}^{1}
\frac{q^2(1-\alpha n_i^m)(1-\alpha n_{i'}^{m'})}
{r(i,m;i',m')} \nonumber \\
     &+& \frac{1}{\beta}\sum_i\sum_{m=0,1}^Z\left[ (1-\alpha n_i^m)
        \ln (1-\alpha n_i^m) + n_i^m\ln n_i^m\right] \;,
\label{2}
\end{eqnarray}
where the average occupation per site is $n_i^m=\langle \sigma_i^m\rangle$
and the constraint, $\sum_i n_i^m=n$, is implicit. 
The optimum upper bound is obtained from the minimization of the 
functional in Eq.~(\ref{2}).  We find,
%---------------------------------
\begin{equation}
\label{4}
n_i^m = \frac{n}{n+\sum_j (1-n_j^m)
\exp[\alpha\xi(\phi_i^m-\phi_j^m)]} \;,
\end{equation}
%---------------------------------
where $\xi=q^2/Dk_BTb$ and
%--------------------------------------------------
\begin{equation}
\label{5}
\phi_i^m=\sum_{j\neq i} \frac{\alpha n_j^m-1}{|i-j|}
+ \sum_j \frac{\alpha n_j^{1-m} -1}{\sqrt{x^2+(i-j)^2}} \;.
\end{equation}
%----------------------------------------
Here $\phi_i^m$ is the reduced electrostatic potential experienced by the
condensed counterion of the $i$'th site of the $m$'th polyion.
These equations can be solved numerically,
producing the positional distribution of the condensed
counterions on the two polyions, Fig 1. We make
the fundamental observation that the two profiles are not
equal.  Thus, the mean-field theory breaks the symmetry
between two polyions!  This is clearly an artifact of
mean-field approximation.  Obviously if the density profiles
would be calculated exactly by an explicit solution
of the partition function, they would be identical.  There
is  no way of breaking the symmetry between two identical 
finite sized
polyions. In the case of exact solution, there would, however, 
exist very strong correlations
between the condensed counterions on the two polyions. These would
provide an important contribution to the total free energy.
Since the mean-field theory does not account for these
correlations, in order to establish an optimum bound, 
it  breaks the symmetry between the
two rods.    

The horizontal component of the force between the two polyions is,
%----------------------------------
\begin{eqnarray}
F_h = \frac{q^2}{Db^2} \sum_{i,j}
      \frac{(1-\alpha n_i^0)(1-\alpha
      n_j^1)}{x^2+|i-j|^2} \; \frac{x}{\sqrt{x^2+|i-j|^2}}\;,
\end{eqnarray}
%-----------------------------------
where $i$ and $j$ correspond to the sites on rods $0$ and
$1$, respectively.
The density profiles $n^0$ and $n^1$ are obtained from
the solution of Eqns.~(\ref{4},\ref{5}).  For short separations
between the polyions the force becomes attractive.  This is 
the result of the symmetry breaking discussed above.  The 
force calculated using the  mean-field theory, Fig. 2, is in quantitative 
agreement with the Monte Carlo simulations and the exact solutions. 

%%%%%%%%%%%%%%%% figure %%%%%%%%%%%%%%%%%%%%%
\begin{figure}[h]
\centerline{\epsfig{file=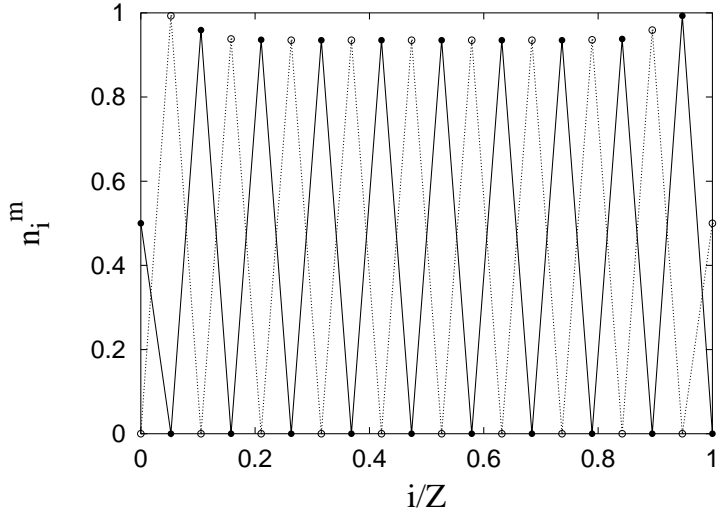,width=7cm,angle=270} }
%\vspace{1cm}
\caption{Density profile for two rods (solid and dashed lines)
for $Z=20$ and $n=9$ divalent counterions, $x=0.4$. Notice the staggered
configurations along the two polyions.}
\label{figure2}
\end{figure}
%%%%%%%%%%%%% end of figure %%%%%%%%%%%%%%%%%                         

%%%%%%%%%%%%%%%% figure %%%%%%%%%%%%%%%%%%%%%
\begin{figure}[ht]
\centerline{\epsfig{file=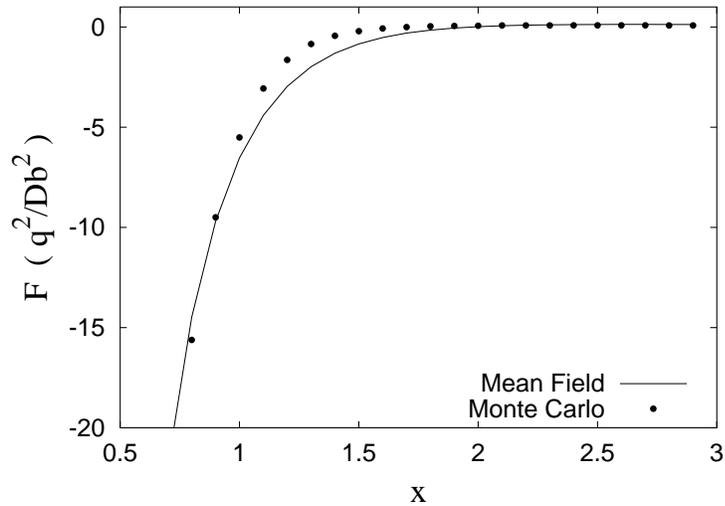,width=7cm,angle=270}}
\caption{The horizontal component of the  force between two like-charged
rods.  $Z=20$ and $n=9$, the net charge on each rod is $-2$.
The points are obtained using
Monte Carlo \cite{Are99}, while the solid line
is the mean field result.}
\label{force}
\end{figure}
%%%%%%%%%%%%% end of figure %%%%%%%%%%%%%%%%%       

The attractive force is short ranged and appears only if the
number of ($\alpha\geq 2$) counterions is larger than a
threshold, $ n = Z/2 \alpha$. For
$\alpha = 1$ the force is always repulsive, which is
in full agreement with the experimental evidence on the absence of
attraction if only monovalent counterions are present \cite{Blo91}.

\end{document}